\documentclass{article}
\usepackage[english]{babel}
\usepackage[latin1]{inputenc}

\usepackage{subfigure}
\usepackage{spconf,amsmath,graphicx}
\usepackage{amsthm}
\usepackage{amssymb}
\usepackage{slashbox}
\usepackage{bbm}
\usepackage{array}

\newcommand{\underoverset}[3]{\overset{#2}{\underset{#1}{#3}}}

\newtheorem{theorem}{Theorem}

\title{Unsupervised diffusion-based LMS for node-specific parameter estimation over wireless sensor networks}

\name{Jorge Plata-Chaves, Mohamad Hasan Bahari, Marc Moonen, Alexander Bertrand \thanks{This work was carried out at the ESAT Laboratory of KU Leuven, in the frame of KU Leuven Research Council CoE PFV/10/002 (OPTEC) and BOF/STG-14-005, Concerted Research Action GOA-MaNet, the Interuniversity Attractive Poles Programme initiated by the Belgian Science Policy Office IUAP P7/23 `Belgian network on stochastic modeling analysis design and optimization of communication systems' (BESTCOM) 2012-2017, Research Project FWO nr. G.0763.12 `Wireless Acoustic Sensor Networks for Extended Auditory Communication', and EU/FP7 project HANDiCAMS. The project HANDiCAMS acknowledges the financial support of the Future and Emerging Technologies (FET) programme within the Seventh Framework Programme for Research of the European Commission, under FET-Open grant number: 323944. The scientific responsibility is assumed by its authors.} } 

 \address{Department of Electrical Engineering-ESAT, STADIUS, KU Leuven, B-3001 Leuven, Belgium \\
 E-mails: \{jplata,  mohamadhasan.bahari, marc.moonen, alexander.bertrand\}@esat.kuleuven.be}

\begin{document}
\ninept
\maketitle
\begin{abstract}

We study a distributed node-specific parameter estimation problem where each node in a wireless sensor network is interested in the simultaneous estimation of different vectors of parameters that can be of local interest, of common interest to a subset of nodes, or of global interest to the whole network. We assume a setting where the nodes do not know which other nodes share the same estimation interests. First, we conduct a theoretical analysis on the asymptotic bias that results in case the nodes blindly process all the local estimates of all their neighbors to solve their own node-specific parameter estimation problem. Next, we propose an unsupervised diffusion-based LMS algorithm that allows each node to obtain unbiased estimates of its node-specific vector of parameters by continuously identifying which of the neighboring local estimates correspond to each of its own estimation tasks. Finally, simulation experiments illustrate the efficiency of the proposed strategy.
\end{abstract}

\section{Introduction}
\label{sec:intro}

In most distributed estimation problems, it is generally assumed that the nodes in a wireless sensor network (WSN) are interested in the same network-wide signal or parameter (e.g.~\cite{mateos2009distributed}\nocite{dimakis2010gossip}\nocite{lopes2007incremental}\nocite{cattivelli2010diffusion}-\cite{chouvardas2011}). However, some applications such as speech enhancement in acoustic sensor networks~\cite{doclo2009reduced}\nocite{bertrand2010distributed}-\cite{plata2015distributed}, beamforming~\cite{bertrand2012lcmv}, or cooperative spectrum sensing in cognitive radio networks~\cite{di2011bio}-\cite{Di_LorenzoTSP} are multi-task oriented. In these cases, special attention is required to more general distributed estimation techniques where the nodes cooperate although they have different but still partially-overlapping estimation interests and their observations may arise from different models. 
 
In the growing literature on node-specific parameter estimation (NSPE) problems over adaptive WSNs, two major groups of works can be identified. The first group assumes that all nodes know a priori the relationship between their estimation tasks and the estimation tasks of their neighbors. Within this category, the aforementioned prior information is leveraged to derive strategies that provide asymptotically unbiased solutions in an NSPE setting where the nodes have both overlapping and arbitrarily different estimation interests~\cite{kekatos2012distributed}-\nocite{platachaves2013a}\nocite{bogdanovic2013aj}\nocite{chendiffusion2014}\cite{platachaves2013aj}. Additionally, this prior information is leveraged by different diffusion-based algorithms that apply different spatial regularizers to let each node solve its estimation task by using the local estimates of neighboring nodes with numerically similar estimation interests~\cite{chen2013multitask}-\cite{nassif2015proximal}. To avoid the bias that results from the combination of local estimates associated with different tasks~\cite{chen2012distributed}-\cite{chen2015multi}, the second group of algorithms implement an inference algorithm together with an adaptive clustering technique that allows the nodes to infer which of their neighbors have the same interest~\cite{chen2015multi}\nocite{zhao2012clustering}\nocite{chen2015multi}\nocite{zhao2015distributed}\nocite{cheneusipco2015}-\cite{khawatmi2015}. However, since the proposed strategies run over diffusion networks where each node is only interested in one vector of parameters, the cooperation is finally limited to nodes that have exactly the same objectives once the clustering technique has converged.

To the best of our knowledge, there are no unsupervised strategies that solve an NSPE problem where the nodes simultaneously estimate parameter vectors of local, common and/or global interest and where there is no prior information about the relationship between the NSPE tasks. Motivated by this fact, we propose an unsupervised diffusion-based LMS for NSPE with combination coefficients determined through a multi-task clustering technique in an adaptive fashion. In this clustering technique, each node solves a hypothesis testing problem to determine which of the local estimates of its neighbors correspond to each of its estimation tasks. Unlike the existing algorithms~\cite{chen2015multi}\nocite{zhao2012clustering}\nocite{chen2015multi}\nocite{zhao2015distributed}\nocite{cheneusipco2015}-\cite{khawatmi2015}, the proposed scheme can yield asymptotically unbiased solutions and allows a beneficial cooperation among the nodes although they have different interests. Indeed, due to the exponential rate of decay for the probabilities of erroneous clustering, as the computer simulations show, the proposed scheme achieves the same steady-state mean square deviation (MSD) as the diffusion-based NSPE LMS (D-NSPE) algorithm that knows a priori the relationship between the node-specific tasks.

\section{Problem formulation}
\label{sec:sec2}

\begin{figure}[t]
\begin{minipage}[b]{1.0\linewidth}
  \centering
  \centerline{\includegraphics[width=0.6\linewidth]{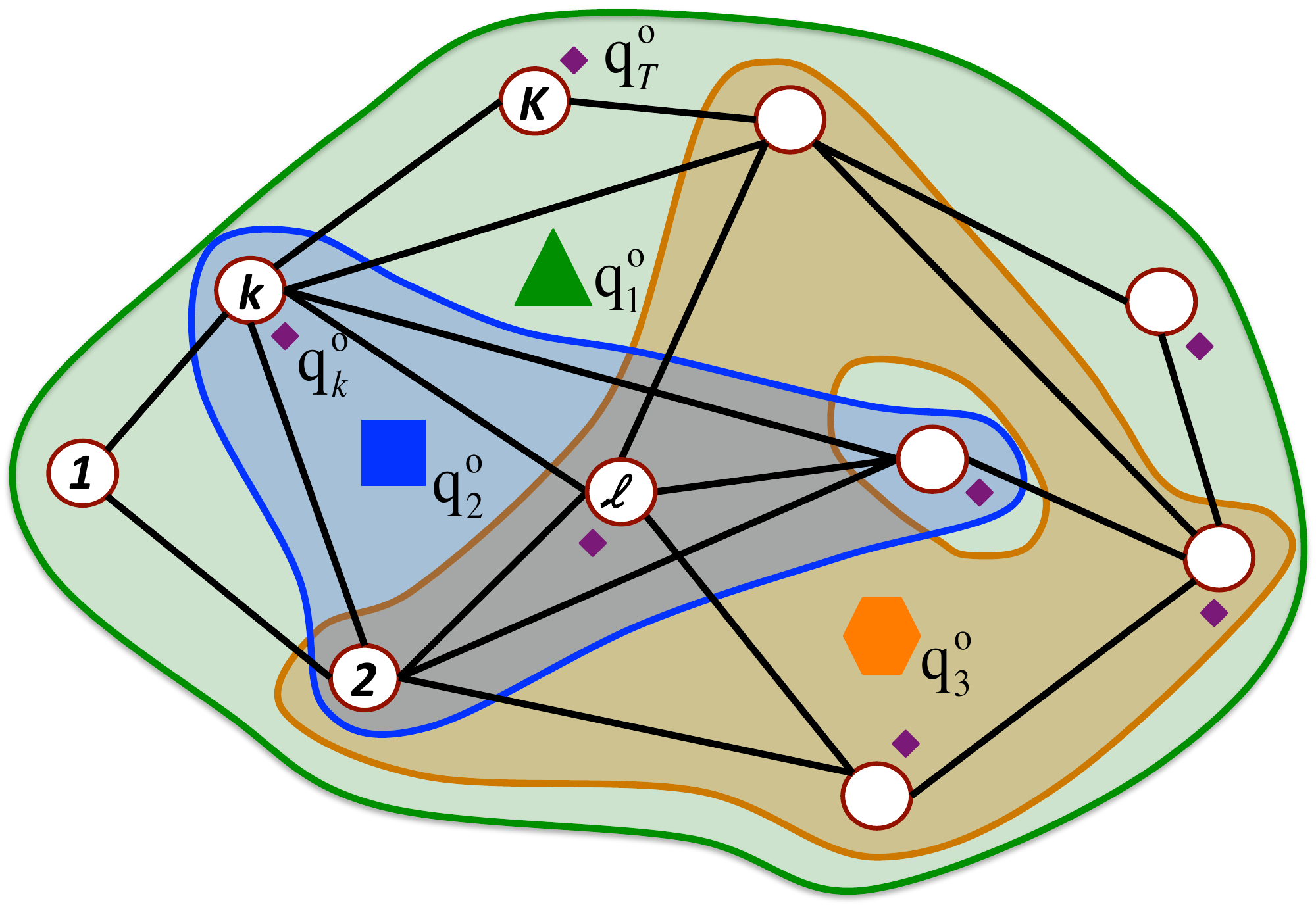}}
  \vspace{-0.15cm}
\end{minipage}
\caption{ A wireless network of $K$ nodes with NSPE interests.}
\label{fig:fig1}
 \vspace{-0.85cm}
\end{figure}

We consider a network consisting of $K$ nodes randomly deployed over some region. Nodes that are able to exchange information in one hop are said to be neighbors. The neighborhood of any particular node $k$, including node $k$ itself, is denoted as $\mathcal{N}_{k}$. To ensure that the network is connected (see Fig.~\ref{fig:fig1}), the neighborhoods are set so that there is a path between any pair of nodes in the network. 

In the considered network, each node $k$, at discrete time $i$, has access to data $ \{d_{k,i}, U_{k,i} \}$, which are realizations related to events that coexist in the monitored area and which follow the relation
\begin{gather}\label{diff:eq1}
\begin{split}
\mathbf{d}_{k,i} &= \mathbf{U}_{k,i} w_k^o + \mathbf{v}_{k,i} 
\end{split}
\end{gather}
where
\begin{itemize}
\item[-] $w_k^o$ equals the vector of dimension $M_k$ that gathers all parameters of interest for node $k$,
\item[-] $\mathbf{v}_{k,i}$ denotes measurement and/or model noise with zero mean and covariance matrix $R_{v_k,i}$ of dimensions $L_k \times L_k$,
\item[-] $\mathbf{d}_{k,i}$ and $\mathbf{U}_{k,i}$ are zero-mean random variables with dimensions $L_k \times 1$ and $L_k \times M_k$, respectively.
\end{itemize}

By processing the data set $\{d_{k,i}, U_{k,i}\}_{k=1}^{K}$, the objective of the network consists in finding the node-specific estimators $\{w_k\}_{k=1}^{K}$ that minimize the following cost function
\begin{gather}\label{diff:eq2}
\begin{split}
J_{\textrm{glob}}(\{w_k\}_{k=1}^{K}) &=\sum_{k=1}^K E \left \{ \Vert \mathbf{d}_{k,i} - \mathbf{U}_{k,i} w_k \Vert^2 \right \}.
\end{split}
\end{gather}
Unlike in papers addressing a single-task estimation problem, e.g.,~\cite{lopes2007incremental}-\cite{cattivelli2010diffusion}, here the parameter estimation interests may vary from node to node in~\eqref{diff:eq1}, i.e., $w_k^o \neq w_\ell^o$ if $k \neq \ell$. Indeed, following the novel node-specific observation model considered in~\cite{bogdanovic2013aj}, \cite{platachaves2013aj}, it is assumed that each node-specific vector $w_k^o$ may consist of a sub-vector of parameters of global interest to the whole network, sub-vectors of parameters of common interest to subsets of nodes including node $k$, and a sub-vector of local parameters for node $k$. In particular, it is considered that
\begin{gather}\label{diff:eq2}
\begin{split}
w_k^o = \mathrm{col} \{ \mathrm{q}^o_t \}_{t \in \mathcal{T}_k}
\end{split}
\end{gather}
where $\mathrm{col}\{ \cdot \}$ denotes a column operator stacking arguments on top of each other and where $\mathrm{q}^o_t$ denotes an $(M_t \times 1)$ vector of parameters associated with the global, common or local estimation task $t$ and where $\mathcal{T}_k$ equals an ordered set of indices $t$ associated with the $n_k = |\mathcal{T}_k|$ vectors $\mathrm{q}_{t}^o$ that are of interest for node $k$. It is noted that $M_k = \sum_{t\in\mathcal{T}_k} M_t$ and that $\mathcal{T}_k \in \mathcal{T}$ where $\mathcal{T}$ is the set of all parameter estimation tasks in the entire network. As a result, the observation model in~\eqref{diff:eq1} can now be rewritten as 
\begin{gather}\label{diff:eq3}
\begin{split}
\mathbf{d}_{k,i} &=\sum_{t \in \mathcal{T}_k} \mathbf{U}_{k_t,i} \mathrm{q}^o_t + \mathbf{v}_{k,i} 
\end{split}
\end{gather}
where 
$\mathbf{U}_{k_t,i}$,  equals a matrix of dimensions $L_k \times M_t$ that consists of the columns of $\mathbf{U}_{k,i}$ associated with $\mathrm{q}^o_t$. From~\eqref{diff:eq3}, also note that the considered NSPE problem can be cast as minimizing
\begin{gather}\label{diff:eq4}
\begin{split}
 \sum_{k=1}^K E \, \Vert \mathbf{d}_{k,i} - \sum_{t \in \mathcal{T}_k} \mathbf{U}_{k_t,i} \mathrm{q}_t \Vert^2 
\end{split}
\end{gather}
with respect to variables $\{\mathrm{q}_{t}\}_{t\in \mathcal{T}}$. 
 
The algorithms derived in~\cite{bogdanovic2013aj} and~\cite{platachaves2013aj} yield unbiased estimates of the node-specific vector of parameters by seeking the minimizer of the cost function in~\eqref{diff:eq4} under an incremental or diffusion mode of cooperation, respectively. However, to do so,  each node should know a priori which of its neighbors share the same parameter estimation interests, i.e., are interested in estimating $\{\mathrm{q}_t^o\}_{t \in \mathcal{T}_k}$. Unfortunately, this prior information is not available in many scenarios. \emph{In this paper, we consider the more challenging problem of deriving a diffusion-based LMS algorithm that is able to solve the NSPE problem stated in~\eqref{diff:eq4} when the nodes do not know a priori which of their neighbors share the same NSPE interests.}  

\section{Performance of the D-NSPE LMS in a setting with unknown relationships between tasks}
\label{sec:sec3}

In this section, in order to motivate the derivation of the proposed algorithm, we briefly analyze the performance of the diffusion-based solution derived in~\cite{platachaves2013aj} when a node does not know which of the local estimates of its neighbors correspond to each of its tasks. For the sake of an easy exposition and without loss of generality, we assume that $M_t=M$ for all $t \in \mathcal{T}$.

In brief, the D-NSPE LMS algorithm~\cite{platachaves2013aj} is able to minimize~\eqref{diff:eq4} by implementing the following recursion at each node $k$:
\begin{gather}\label{diff:eq5}
\begin{split}
\left \{
\begin{array}{ll}
\textrm{\textbf{Adaptation step.} For each task } t \in \mathcal{T}_k \textrm{ execute} \\
\boldsymbol{\psi}_{k,t}^{(i)} =   \boldsymbol{\phi}_{k,t}^{(i-1)}  + \mu_k \, \mathbf{U}_{k_t,i}^H \left [ \mathbf{d}_{k,i} - \sum_{p \in \mathcal{T}_k}\mathbf{U}_{k_p,i} \boldsymbol{\phi}_{k,p}^{(i-1)} \right ]  \\
\textrm{\textbf{Combination step.} For each task } t \in \mathcal{T}_k \textrm{ execute} \\
 \boldsymbol{\phi}_{k,t}^{(i)} = \sum_{\ell \in \mathcal{N}_k}   \sum_{p\in \mathcal{T}_\ell}  c_{k\ell,tp}(i)  \boldsymbol{\psi}_{\ell,p}^{(i)}.
\end{array}
\right .
\end{split}
\end{gather}
where $\boldsymbol{\phi}_{k,t}^{(i)}$ is the local estimate of $\mathrm{q}_t^o$ at node $k$ and time instant $i$ and $\{c_{k\ell,tp}(i)\}$ are time-varying convex combination coefficients that satisfy
\begin{gather}\label{diff:eq6}
\begin{split}
\left \{
\begin{array}{ll}
 c_{k\ell,tp}(i) > 0 & \textrm{if } \ell \in \mathcal{N}_k \cap \mathcal{C}_t \textrm{ and } p = t,\\
 c_{k\ell,tp}(i) = 0 & \textrm{otherwise}.
 \end{array}
 \right .
\end{split}
\end{gather}
and
\begin{gather}\label{diff:eq7}
\begin{split}
\sum_{\ell \in \mathcal{N}_k}   \sum_{p\in \mathcal{T}_\ell} c_{k\ell,tp}(i) = 1
\end{split}
\end{gather}
with $\mathcal{C}_t$ denoting the set of nodes interested in estimating $\mathrm{q}_t^o$. Note that there are several policies to select the combination coefficients. On the one hand, the combination rule can be static, e.g., as in the uniform, Metropolis or relative-degree rule~\cite{sayed2012diffusion}. On the other hand, the coefficients can be adapted over time (e.g, see~\cite{zhao2012clustering}). 

Independently of the selected combination policy, from~\eqref{diff:eq5} and~\eqref{diff:eq6}, it can be noticed that the combination step associated with the estimation of $\mathrm{q}_t^o$ at node $k$ can only process local estimates of the same vector of parameters, i.e., $p=t$, transmitted by neighboring nodes, i.e., $\ell \in \mathcal{N}_k$. This particular constraint on the set of possible combination policies ensures asymptotical unbiasedness~\cite{platachaves2013aj}, i.e., $\mathrm{lim}_{i \to \infty} E \{ \boldsymbol{\tilde{q}}_{k,t}^{(i)} \} = 0$ with $ \boldsymbol{\tilde{\mathrm{q}}}_{k,t}^{(i)} = \mathrm{q}_t^o - \mathrm{col} \{ \boldsymbol{\phi}_{k,t}^{(i)}\}_{t\in \mathcal{T}_k}$ and $w_k^o$ defined in~\eqref{diff:eq2}. However, it requires each node to know a priori which of the local estimates exchanged by its neighbors correspond to each of its parameter estimation tasks. 

When the nodes do not know a priori the relationship between the tasks, each node could implement a stand-alone LMS to solve its NSPE problem, which is equivalent to the implementation of~\eqref{diff:eq5}-\eqref{diff:eq7} with $c_{kk,tt}(i) = 1$. Although this approach would again allow to find unbiased estimates of the node-specific vector of parameters that minimize~\eqref{diff:eq4}, the nodes cannot take advantage of the well-known benefits provided by the cooperation. Alternatively, each node could blindly fuse all the local estimates exchanged by all its neighbors, which yields an implementation of~\eqref{diff:eq5} where the convex combination coefficients satisfy~\eqref{diff:eq7} and
\begin{gather}\label{diff:eq9}
\begin{split}
\left \{
\begin{array}{ll}
 c_{k\ell,tp}(i) =  \alpha_{k\ell,tp} > 0 & \textrm{if } \ell \in \mathcal{N}_k,\\ 
 c_{k\ell,tp}(i) = 0 & \textrm{otherwise}.
 \end{array}
 \right .
\end{split}
\end{gather}
with $\alpha_{k\ell,tp}$ equal to a positive constant for all $k \in \{1,2,\ldots,K\}$, $\ell \in \mathcal{N}_k$, $t\in \mathcal{T}_k$ and $p \in \mathcal{T}_\ell$. However, under this approach the estimates at the nodes will be biased. In particular, assuming that 
\begin{itemize}
\item[A1)]  $\mathbf{v}_{k,i}$ is temporally and spatially white noise that is independent of  $\mathbf{U}_{\ell,i'}$ for all $\ell$ and $i'$, with $k,\ell \in \{1,2,\ldots,K\}$. 
\item[A2)]  $\mathbf{U}_{k,i}$ is temporally stationary, white and spatially independent with $R_{U_k}=E\{ \mathbf{U}_{k,i} \mathbf{U}_{k,i}^{H} \}$; 
\item[A3)] $\mathbf{U}_{k_t,i}$ and $\mathbf{U}_{k_p,i}$ are independent for all $k \in \{1,2, \ldots, K\}$ and $t \neq p$,
\end{itemize}
the asymptotic bias of the estimates resulting from this cooperative approach is given by the following theorem (the proof is omitted due to space constraints). 

\begin{theorem}\label{teo:teo1} 
For any initial conditions and under assumptions A1-A3, if the positive step-size of each node satisfies 
\begin{gather}\label{diff:eq10}
\begin{split}
\mu_k < 2/\lambda_{\mathrm{max}}( \{R_{U_{k_{t}}}\}_{t \in \mathcal{T}_k}), 
\end{split}
\end{gather}
then the estimates generated by the D-NSPE LMS algorithm summarized in~\eqref{diff:eq5} converge in the mean when the convex combination coefficients satisfy~\eqref{diff:eq7} and~\eqref{diff:eq9}. Furthermore, the estimation bias in the steady-state tends to
\begin{gather}\label{diff:eq11}
\begin{split}
E \{ \boldsymbol{\tilde{\mathrm{q}}}^{(\infty)} \} = \big [I-\breve{C} [I - M D] \big]^{-1} [I-\breve{C}] \, \mathrm{q}^{o} 
\end{split}
\end{gather}
with $\boldsymbol{\tilde{\mathrm{q}}}^{(i)} = \mathrm{col}\{\mathrm{col}\{  \boldsymbol{\tilde{\mathrm{q}}}_{k,t}^{(i)} \}_{t \in \mathcal{T}_k}\}_{k=1}^{K}$, $\mathrm{q}^{o} = \mathrm{col}\{\mathrm{col} \{\mathrm{q}^{o}_{t}\}_{t \in \mathcal{T}_k} \}_{k=1}^{K}$ $D=\mathrm{diag}\{R_{U_k} \}_{k=1}^{K}$, $M=\mathrm{diag}\{\mu_k I_{M_k} \}_{k=1}^{K}$ 
and $\breve{C} = C \otimes I_M$ where
\begin{gather}\label{diff:eq12}
\begin{split}
C=\left [ c_{1,\mathcal{T}_1(1)} \, \cdots \, c_{1,\mathcal{T}_1(n_1)} \, \, c_{2,\mathcal{T}_2(1)} \, \cdots \, c_{K,\mathcal{T}_K(n_K)} \right ]^T
\end{split}
\end{gather}
and $c_{k,\mathcal{T}_k(t)} = \mathrm{col} \big \{ \{\mathrm{col}\{c_{k\ell,\mathcal{T}_k(t)p}\}_{p \in \mathcal{T}_{\ell}} \}_{\ell=1}^{K}  \big \}$.
\end{theorem}
 
From Theorem~\ref{teo:teo1}, we can deduce that the steady-state MSD in the estimation of $\mathrm{q}_t^{o}$ at node $k$, i.e., $\mathrm{lim}_{i \to \infty} \lVert \tilde{\mathbf{q}}_{k,t}^{i} \rVert^{2}$, can be very large when node $k$ estimates $\mathrm{q}_t^o$ by using a blind convex combination of all the local estimates of all its neighbors. As a result, the attained performance might be worse than the one achieved by the non-cooperative scheme. Note that~\eqref{diff:eq11} reduces to zero if $c_{kk,tt}=1$ for all $k$ and $t \in \mathcal{T}_k$, i.e., the non-cooperative case is indeed unbiased. Next, we will propose a scheme that combines the D-NSPE LMS algorithm and an adaptive multi-task clustering technique to avoid such a performance degradation and still leverage the cooperation among nodes interested in estimating different but overlapping vectors of parameters simultaneously. 

\section{Unsupervised diffusion-based LMS for NSPE} 
\label{sec:sec4}

Since the nodes do not know a priori the relationship between the NSPE tasks, they can initially focus on solving the NSPE problem of Section~\ref{sec:sec2} by implementing the non-cooperative strategy. In particular, each agent can implement~\eqref{diff:eq5}-\eqref{diff:eq7} with $c_{kk,tt}(i)=1$ for all $k \in \{1,2, \ldots, K\}$ and $t \in \mathcal{T}_k$, which corresponds to the following stand-alone LMS-type adaptation step
\begin{gather}\label{diff:eq15}
\begin{split}
\boldsymbol{\varsigma}_{k,t}^{(i)} =   \boldsymbol{\varsigma}_{k,t}^{(i-1)}  + \mu_k \, \mathbf{U}_{k_t,i}^H \Big [ \mathbf{d}_{k,i} - \sum_{p \in \mathcal{T}_k}\mathbf{U}_{k_p,i} \boldsymbol{\varsigma}_{k,p}^{(i-1)} \Big ] 
\end{split}
\end{gather}
where $\boldsymbol{\varsigma}_{k,t}^{(i)}$ denotes the estimate of $\mathrm{q}_{t}^{o}$ resulting from the non-cooperative LMS performed by node $k$ at time instant $i$ with $t \in \mathcal{T}_k$. 

Following similar arguments to~\cite{zhao2015distributed}, for sufficiently small step sizes, i.e., $\mu_k \ll 1$, and after sufficient iterations, i.e., $i \to \infty$, it can be shown that the difference between $\boldsymbol{\varsigma}_{k,t}^{(i)}$ and $\boldsymbol{\varsigma}_{\ell,p}^{(i)}$ follows a Gaussian distribution:
\begin{gather}\label{diff:eq16}
\begin{split}
\boldsymbol{\varsigma}_{k,t}^{(i)}-\boldsymbol{\varsigma}_{\ell,p}^{(i)} \sim \mathbb{N}(\mathrm{q}_{t}^{o}-\mathrm{q}_{p}^{o},\mu_{\mathrm{max}}\Delta_{k\ell,tp})
\end{split}
\end{gather}
where $\mu_{\mathrm{max}}=\underset{k}{\mathrm{max}} \, \mu_k$ and $\Delta_{k\ell,tp}$ is an $M\times M$ symmetric, positive semi-definite matrix with $k,\ell \in \{1,2,\ldots,K\}$, $t \in \mathcal{T}_k$ and $p \in \mathcal{T}_\ell$. If both estimates are associated with the same task, which is denoted as $\mathrm{q}_t^o = \mathrm{q}_p^o$ from now on, from~\eqref{diff:eq16} note that with high probability $\lVert \boldsymbol{\varsigma}_{k,t}^{(i)}-\boldsymbol{\varsigma}_{\ell,p}^{(i)} \rVert^{2} = O(\mu_{\mathrm{max}})$. On the contrary, if the local estimates are associated with different tasks, which will be denoted as $\mathrm{q}_t^o \neq \mathrm{q}_p^o$, then $\lVert \boldsymbol{\varsigma}_{k,t}^{(i)}-\boldsymbol{\varsigma}_{\ell,p}^{(i)} \rVert^{2} = O(1)$ will hold with high probability. Thus, if node $\ell$ transmits $\boldsymbol{\varsigma}_{\ell,p}^{(i)}$ to node $k$ with $\ell \in \mathcal{N}_k \setminus \{k\}$ and $p \in \mathcal{T}_{\ell}$, node $k$ can perform a hypothesis test to determine whether the local estimate $\boldsymbol{\varsigma}_{\ell,p}^{(i)}$ at node $\ell$ corresponds to the estimation of the vector of parameters $\mathbf{q}_{t}^{o}$ with $t \in \mathcal{T}_{k}$:
\begin{gather}\label{diff:eq17}
\begin{split}
\lVert \boldsymbol{\varsigma}_{k,t}^{(i)}-\boldsymbol{\varsigma}_{\ell,p}^{(i)} \rVert^{2} \underoverset{\mathbb{H}_1}{\mathbb{H}_0}{\gtrless} \tau_{k\ell,tp}
\end{split}
\end{gather}
where $\mathbb{H}_0$ equals the hypothesis $\mathrm{q}_t^o = \mathrm{q}_p^o$, $\mathbb{H}_1$ denotes the hypothesis $\mathrm{q}_t^o \neq \mathrm{q}_p^o$ and $\tau_{k\ell,tp}$ is a predefined threshold.

\begin{figure}[t]
\centering 
\subfigure[]{
\includegraphics[width=0.33\textwidth]{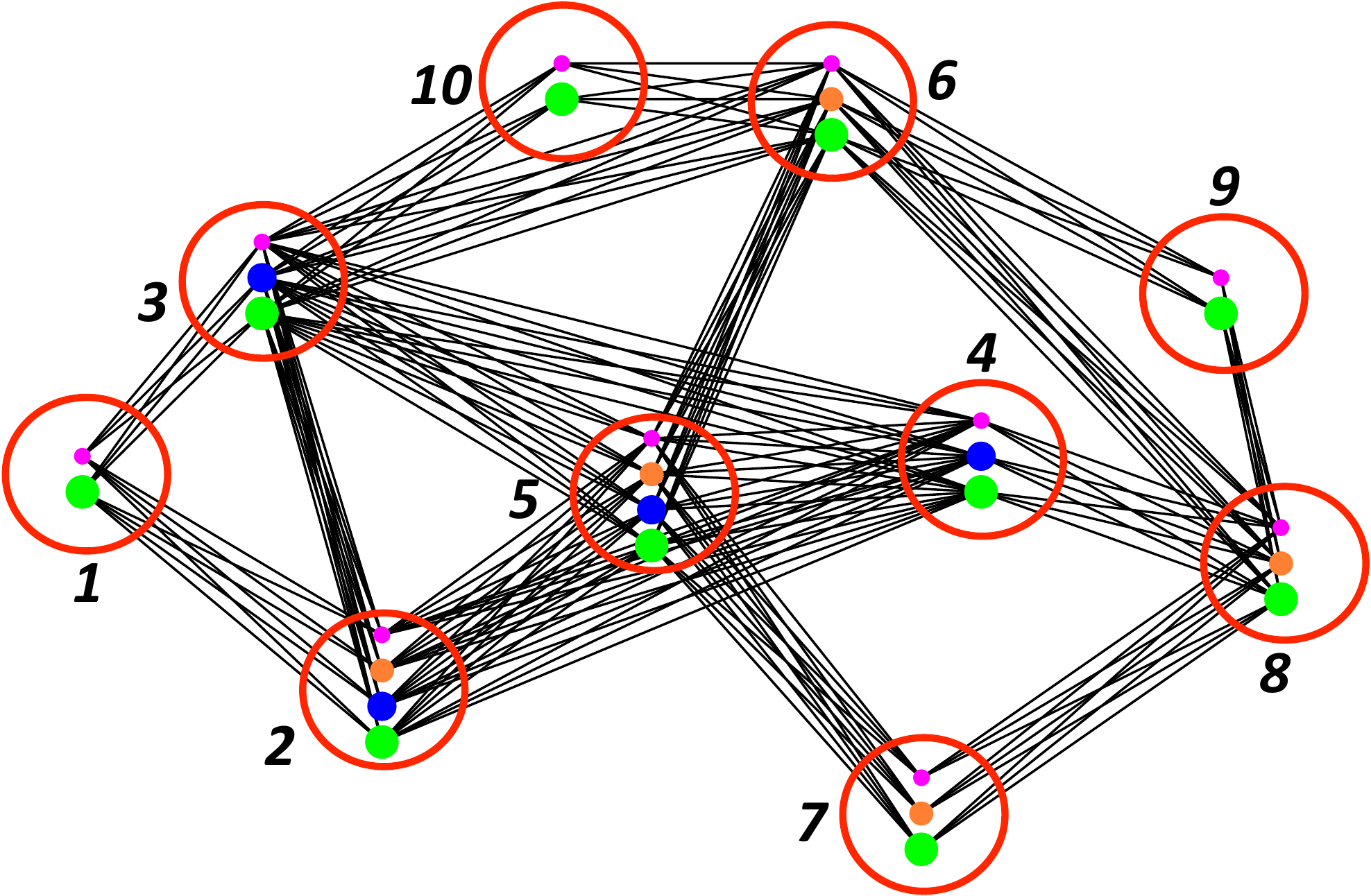}
\label{fig:fig4a}
} 
\subfigure[]{
\includegraphics[width=0.33\textwidth]{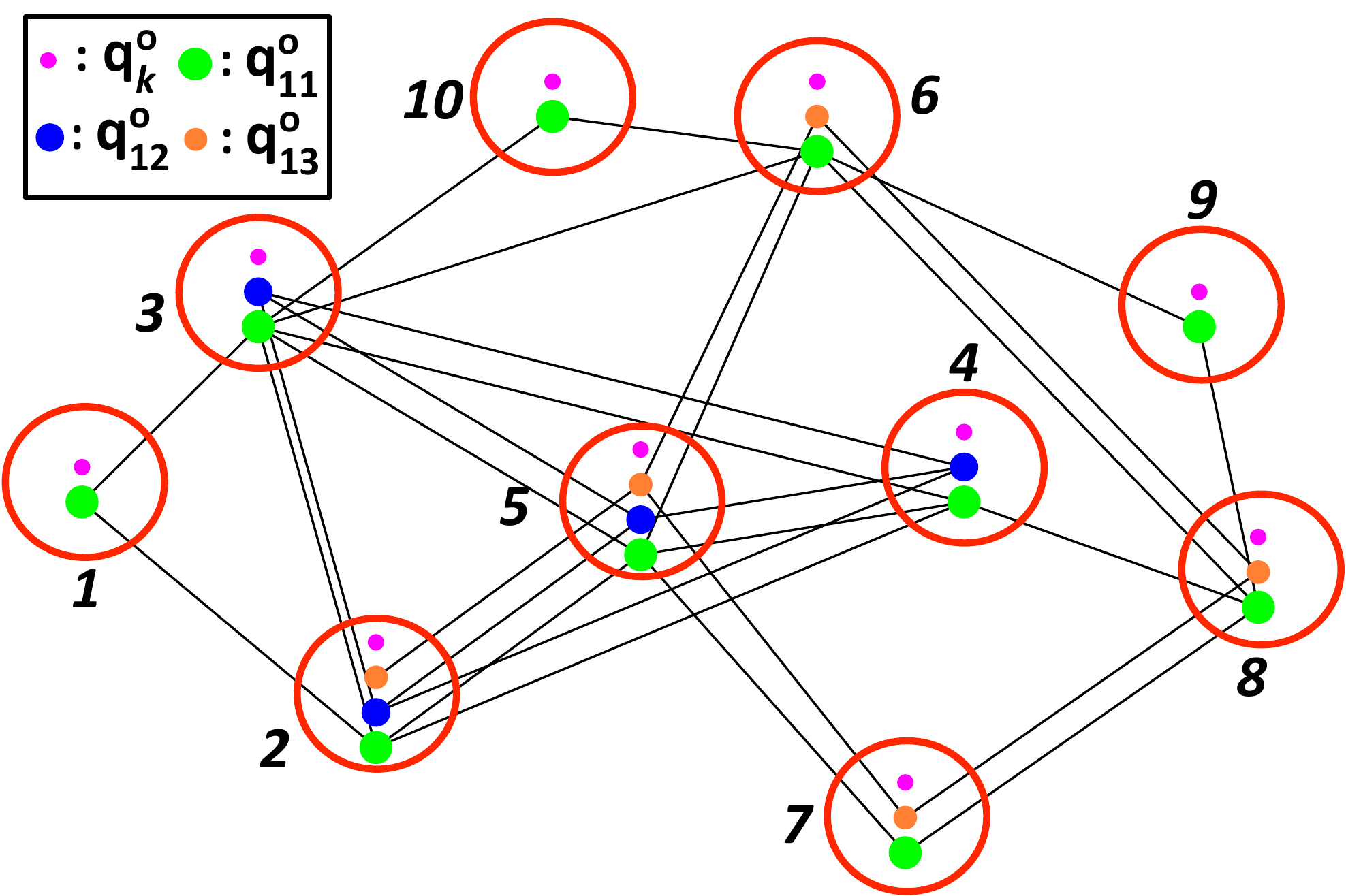}
\label{fig:fig4b}
}
\caption{Topology with all the initial cooperation links~\subref{fig:fig4a} and the resulting cooperation links after multi-task clustering~\subref{fig:fig4b}.} 
\label{fig:fig4}
\end{figure}

A similar reasoning to~\cite{zhao2015distributed} shows that the probabilities of false alarm and misdetection decay at exponential rates, i.e., 
\begin{gather}\label{diff:eq18}
\begin{split}
\mathbb{P}\big[ \lVert \boldsymbol{\varsigma}_{k,t}^{(i)}-\boldsymbol{\varsigma}_{\ell,p}^{(i)} \rVert^{2} > \tau_{k\ell,tp} \mid  \mathrm{q}_t^o = \mathrm{q}_p^o  \big]  \leq O(e^{-c_1/\mu_{\mathrm{max}}})
\end{split}
\end{gather}
\begin{gather}\label{diff:eq19}
\begin{split}
\mathbb{P}\big[ \lVert \boldsymbol{\varsigma}_{k,t}^{(i)}-\boldsymbol{\varsigma}_{\ell,p}^{(i)} \rVert^{2} < \tau_{k\ell,tp} \mid \mathrm{q}_t^o \neq \mathrm{q}_p^o  \big]  \leq O(e^{-c_2/\mu_{\mathrm{max}}})
\end{split}
\end{gather}
where $c_1,c_2 > 0$ denotes some positive constants and $\tau_{k\ell,tp} \in (0,d_{k\ell,tp})$ with $d_{k\ell,tp} = \lVert q_t^{o} - q_p^{o} \rVert^2$. Thus, if $\tau_{k\ell,tp}$ and $\mu_{\mathrm{max}}$ are sufficiently small, as $i$ approaches infinity, node $k$ can know with high probability if a local estimate $\boldsymbol{\varsigma}_{\ell,p}^{(i)} $ at node $\ell$ is associated with the estimation of $q_t^{o}$. This information can be used by node $k$ to simultaneously cluster the local estimates of the nodes according to the task that they aim to solve. In particular, from the following set 
\begin{gather}\label{diff:eq20}
\begin{split}
\boldsymbol{\mathcal{N}}_{k,t}(i)=& 
  \left \{ (\ell,p): \ell \in \mathcal{N}_k, \,  p \in \mathcal{T}_\ell,  \lVert \boldsymbol{\varsigma}_{k,t}^{(i)}-\boldsymbol{\varsigma}_{\ell,p}^{(i)} \rVert^{2} < \tau_{k\ell,tp}  \right \}
\end{split}
\end{gather}
node $k$ can dynamically infer the set of indices $\ell \in \mathcal{N}_\ell$ and $p \in \mathcal{T}_\ell$ for which $c_{k\ell,tp}(i)> 0$ should be verified. At the same time, these task-specific neighborhoods can be used by each node $k$ to perform the following diffusion-based NSPE strategy for each task $t \in \mathcal{T}_k$
\begin{gather}\label{diff:eq21}
\begin{split}
\left \{
\begin{array}{ll}
\boldsymbol{\psi}_{k,t}^{(i)} =   \boldsymbol{\phi}_{k,t}^{(i-1)}  + \mu_k \, \mathbf{U}_{k_t,i}^H \left [ \mathbf{d}_{k,i} - \sum_{p \in \mathcal{T}_k}\mathbf{U}_{k_p,i} \boldsymbol{\phi}_{k,p}^{(i-1)} \right ]  \\
 \boldsymbol{\phi}_{k,t}^{(i)} = \sum_{(\ell,p) \in \boldsymbol{\mathcal{N}}_{k,t}(i-1)}  c_{k\ell,tp}(i-1)  \boldsymbol{\psi}_{\ell,p}^{(i)}.
\end{array}
\right .
\end{split}
\end{gather}
where $\sum_{(\ell,p) \in \boldsymbol{\mathcal{N}}_{k,t}(i-1)}  c_{k\ell,tp}(i-1) =1$.

This leads to the following algorithm:\\
\rule{\linewidth}{0.5mm} \\[-0.5mm]
\textbf{Unsupervised Diffusion-based LMS for NSPE (UD-NSPE)}\\[-2mm]
\rule{\linewidth}{0.5mm}
\begin{itemize}
\item Start with any initial guesses $\varsigma_{k,t}^{(0)}$, $\phi_{k,t}^{(0)}$ and $\boldsymbol{\mathcal{N}}_{k,t}(-1)=\{(k,t)\}$ for all $k \in \{1,2,\ldots,K\}$ and $t \in \mathcal{T}_k$.
\item At each time $i$ and each node $k \in \{1,2,\ldots,K\}$:
\begin{itemize}
\item[1.] Update $\boldsymbol{\varsigma}_{k,t}^{(i)}$ by executing~\eqref{diff:eq15} for each $t \in \mathcal{T}_k$.
\item[2.] Update $ \boldsymbol{\phi}_{k,t}^{(i)}$ by executing recursion~\eqref{diff:eq21} over the set $\boldsymbol{\mathcal{N}}_{k,t}(i-1)$ for each $t \in \mathcal{T}_k$.
\item[3.] Update the set $\boldsymbol{\mathcal{N}}_{k,t}(i)$ for each $t \in \mathcal{T}_k$ by using~\eqref{diff:eq20} with $\{\varsigma_{\ell,p}^{(i)}; \ell \in \mathcal{N}_k, p \in \mathcal{T}_\ell \}$ from Step 1.
\end{itemize}
\end{itemize}
\vspace{-2mm}
\rule{\linewidth}{0.5mm}\\[-1mm]

Based on the multi-task clustering information resulting from~\eqref{diff:eq15} and~\eqref{diff:eq20}, at each time instant $i$ the proposed UD-NSPE LMS is able to leverage the cooperation among nodes with different but overlapping estimation interests. Despite this fact, from~\eqref{diff:eq19}, note that UD-NSPE LMS still yields asymptotically unbiased estimates if the $\lim_{i \to \infty} \mu_k = 0$ for all $k \in \{1,2,\ldots,K\}$. 

\section{Simulations}\label{sec:sec5}

\begin{figure}[t]
 \vspace{-0.25cm}
\begin{minipage}[b]{1.0\linewidth}
  \centering
  \centerline{\includegraphics[width=1.12\linewidth]{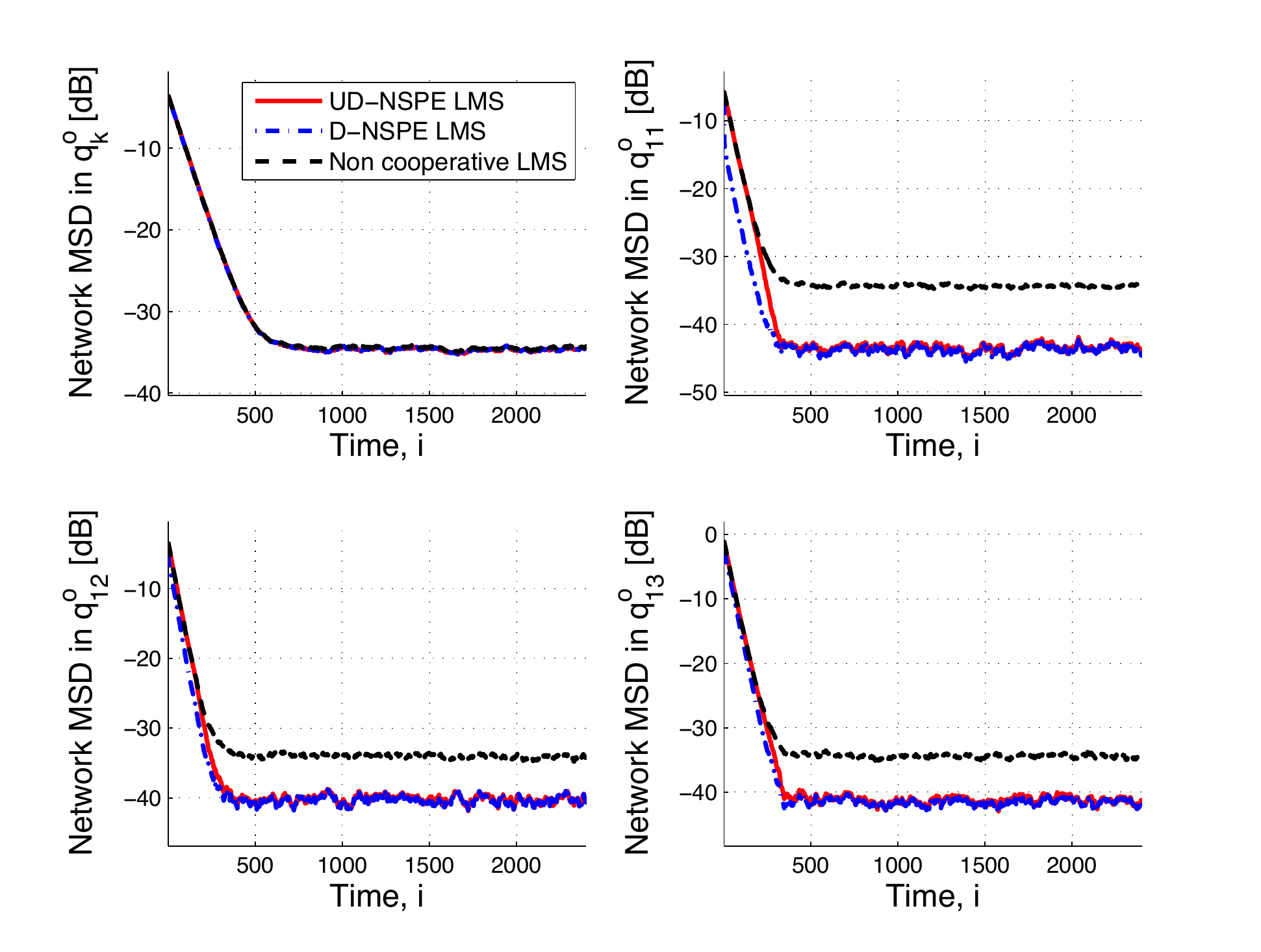}} 
  \vspace{-0.15cm}
\end{minipage}
\caption{Temporal evolution of the network MSD for the estimation of the different vectors of parameters.} 
\label{fig:fig3}
 \vspace{-0.5cm}
\end{figure}

To illustrate the effectiveness of the proposed algorithm, we consider a network formed by $K=10$ nodes whose initial topology with all possible cooperation links is shown in Fig.~\ref{fig:fig4a}. The red circles denote the nodes and the colored balls inside represent different tasks within each node. In the considered network, each node $k$ is interested in estimating a vector of global parameters $\mathrm{q}^o_{11} \in \mathbb{R}^3$ and a vector of local parameters $\mathrm{q}_k^o \in \mathbb{R}^3$ with $k \in \{1,\ldots,K\}$, denoted with green and magenta balls, respectively. Additionally, two different vectors of common parameters coexist, i.e., $\mathrm{q}_{12}^{o} \in \mathbb{R}^{3}$ and $\mathrm{q}_{13}^{o} \in \mathbb{R}^{3}$, represented by blue and orange balls, respectively. Each entry of the global, common or local vector of parameters is randomly drawn from an uniform distribution defined in the interval (0,1). Moreover, the data observed by node $k$ follows the observation model given in~\eqref{diff:eq1} with $L_k=1$.  Both the the background noise $v_{k,i}$ and the regressors $\mathbf{u}_{k,i}$ are independently drawn from a Gaussian distribution that is spatially and temporally independent. In particular, $v_{k,i}$ follows a Gaussian distribution with zero mean and variance $\sigma_{v_k}^2=10^{-3}$ for all $k$. Similarly, the regressors $\mathbf{u}_{k,i}$ are zero mean $(1 \times 3 n_k)$ random vectors governed by a Gaussian distribution with zero mean and covariance matrix $\mathbf{R}_{\mathbf{u}_{k,i}}=\sigma_{\mathbf{u}_{k}}^2\mathbf{I}_{3 n_k}$. The variance $\sigma_{\mathbf{u}_{k}}^2$ is randomly chosen in $(0,1)$ so that the Signal-to-Noise-Ratio (SNR) at each node ranges from 10 dB to 20 dB. Furthermore, each step-size $\mu_k$ is set equal to  $4 \cdot 10^{-3}$ and a uniform combination policy has been selected to generate the combination coefficients. 

Fig.~\ref{fig:fig4b} shows the estimated cooperation links in the steady-state for one of the experiments. Note that cooperation links between local estimates of the same vector of parameters at different neighboring nodes remain active. On the contrary, cooperation links between local estimates of different vectors of parameters are dropped. For the UD-NSPE algorithm, the D-NSPE algorithm derived in~\cite{platachaves2013aj} and the non-cooperative LMS, Fig.~\ref{fig:fig3} plots the learning behaviour in terms of the network MSD associated with the estimation of the vectors of global, common and local parameters. To generate each plot, we have averaged the results over 100 randomly initialized independent experiments. Since the multi-task clustering technique correctly determines the links between local estimates of the same vector of parameters at different neighboring nodes, the UD-NSPE algorithm outperforms the non-cooperative LMS and achieves the same steady-state network MSD as the D-NSPE algorithm. 
 
\section{Conclusion}\label{sec:con}

We have considered an NSPE problem where the nodes simultaneously estimate vectors of local, common and global interest in a setting where the nodes do not know a priori which of the local estimates of their neighbors correspond to which estimation task. To solve this problem, we have presented an algorithm based on a diffusion-based NSPE LMS and a multi-task clustering technique that lets each node infer which of the local estimates of its neighbors correspond to each of its own estimation tasks. Unlike the existing schemes, the proposed algorithm can yield unbiased estimates for the NSPE problem while still leveraging the cooperation among nodes with different interests. Finally, the effectiveness of the proposed algorithm has been illustrated through computer simulations.

\bibliographystyle{IEEEbib}
\bibliography{references}

\begin{thebibliography}{10}

\bibitem{mateos2009distributed}
G.~Mateos, I.~D. Schizas, and G.~B. Giannakis,
\newblock ``Distributed recursive least-squares for consensus-based in-network
  adaptive estimation,''
\newblock {\em IEEE Transactions on Signal Processing}, vol. 57, no. 11, pp.
  4583--4588, 2009.

\bibitem{dimakis2010gossip}
A.~G. Dimakis, S.~Kar, J.~M.~F. Moura, M.~G. Rabbat, and A.~Scaglione,
\newblock ``Gossip algorithms for distributed signal processing,''
\newblock {\em Proceedings of the IEEE}, vol. 98, no. 11, pp. 1847--1864, 2010.

\bibitem{lopes2007incremental}
C.~G. Lopes and A.~H. Sayed,
\newblock ``Incremental adaptive strategies over distributed networks,''
\newblock {\em IEEE Transactions on Signal Processing}, vol. 55, no. 8, pp.
  4064--4077, 2007.

\bibitem{cattivelli2010diffusion}
F.~S. Cattivelli and A.~H. Sayed,
\newblock ``Diffusion {L}{M}{S} strategies for distributed estimation,''
\newblock {\em IEEE Transactions on Signal Processing}, vol. 58, no. 3, pp.
  1035--1048, 2010.

\bibitem{chouvardas2011}
S.~Chouvardas, K.~Slavakis, and S.~Theodoridis,
\newblock ``Adaptive robust distributed learning in diffusion sensor
  networks,''
\newblock {\em IEEE Transactions on Signal Processing}, vol. 59, no. 10, pp.
  4692--4707, 2011.

\bibitem{doclo2009reduced}
S.~Doclo, M.~Moonen, T.~Van~den Bogaert, and J.~Wouters,
\newblock ``Reduced-bandwidth and distributed mwf-based noise reduction
  algorithms for binaural hearing aids,''
\newblock {\em IEEE Transactions on Audio, Speech, and Language Processing},
  vol. 17, no. 1, pp. 38--51, 2009.

\bibitem{bertrand2010distributed}
A.~Bertrand and M.~Moonen,
\newblock ``Distributed adaptive node-specific signal estimation in fully
  connected sensor networks - part {I}: Sequential node updating,''
\newblock {\em IEEE Transactions on Signal Processing}, vol. 58, no. 10, pp.
  5277--5291, 2010.

\bibitem{plata2015distributed}
J.~Plata-Chaves, A.~Bertrand, and M.~Moonen,
\newblock ``Distributed signal estimation in a wireless sensor network with
  partially-overlapping node-specific interests or source observability,''
\newblock in {\em IEEE 40th International Conference on Acoustics, Speech and
  Signal Processing, 2015. ICASSP 2015}, 2015.

\bibitem{bertrand2012lcmv}
A.~Bertrand and M.~Moonen,
\newblock ``Distributed node-specific {L}{C}{M}{V} beamforming in wireless
  sensor networks,''
\newblock {\em IEEE Transactions on Signal Processing}, vol. 60, no. 1, pp.
  233--246, 2012.

\bibitem{di2011bio}
P.~Di~Lorenzo, S.~Barbarossa, and A.~H. Sayed,
\newblock ``{Bio-inspired swarming for dynamic radio access based on diffusion
  adaptation},''
\newblock in {\em IEEE 19th European Signal Conference, 2011. EUSIPCO 2011},
  2012, pp. 1--6.

\bibitem{Di_LorenzoTSP}
P.~Di~Lorenzo, S.~Barbarossa, and A.~H. Sayed,
\newblock ``Bio-inspired decentralized radio access based on swarming
  mechanisms over adaptive networks,''
\newblock {\em IEEE Transactions on Signal Processing}, vol. 61, no. 12, pp.
  3183--3197, 2013.

\bibitem{kekatos2012distributed}
V.~Kekatos and G.~B. Giannakis,
\newblock ``Distributed robust power system state estimation,''
\newblock {\em IEEE Transactions on Power Systems}, vol. 28, no. 2, pp.
  1617--1626, 2013.

\bibitem{platachaves2013a}
J.~Plata-Chaves, N.~Bogdanovic, and K.~Berberidis,
\newblock ``{Distributed incremental-based {R}{L}{S} for node-specific
  parameter estimation over adaptive networks},''
\newblock in {\em IEEE 21st European Signal Conference, 2013. EUSIPCO 2013},
  2013.

\bibitem{bogdanovic2013aj}
N.~Bogdanovic, J.~Plata-Chaves, and K.~Berberidis,
\newblock ``{Distributed incremental-based {L}{M}{S} for node-specific adaptive
  parameter estimation},''
\newblock {\em IEEE Transactions on Signal Processing}, vol. 62, no. 20, pp.
  5382--5397, 2014.

\bibitem{chendiffusion2014}
J.~Chen, C.~Richard, A.~O. Hero~III, and A.~H. Sayed,
\newblock ``Diffusion {L}{M}{S} for multitask problems with overlapping
  hypothesis subspaces,''
\newblock in {\em IEEE International Workshop on Machine Learning for Signal
  Processing, MLSP 2014}, 2014.

\bibitem{platachaves2013aj}
J.~Plata-Chaves, N.~Bogdanovic, and K.~Berberidis,
\newblock ``{Distributed diffusion-based {L}{M}{S} for node-specific parameter
  estimation over adaptive networks},''
\newblock {\em IEEE Transactions on Signal Processing}, vol. 13, no. 63, pp.
  3448--3460, 2015.

\bibitem{chen2013multitask}
J.~Chen, C.~Richard, and A.~H. Sayed,
\newblock ``Multitask diffusion adaptation over networks,''
\newblock {\em Submitted to Transactions on Signal Processing}, 2013 [Online].
  Available: http://arxiv.org/abs/1311.4894.

\bibitem{nassif2015proximal}
R.~Nassif, C.~Richard, A.~Ferrari, and Ali~H Sayed,
\newblock ``Proximal multitask learning over networks with sparsity-inducing
  coregularization,''
\newblock {\em Submitted to IEEE Transactions on Signal Processing}, 2015
  [Online]. Available: http://arxiv.org/abs/1509.01360.

\bibitem{chen2012distributed}
J.~Chen and A.~H. Sayed,
\newblock ``{Distributed Pareto-optimal solutions via diffusion adaptation},''
\newblock in {\em IEEE Statistical Signal Processing Workshop, 2012. SSP
  2012.}, 2012, pp. 648--651.

\bibitem{chen2015multi}
J.~Chen, C.~Richard, and A.~H. Sayed,
\newblock ``Diffusion {L}{M}{S} over multitask networks,''
\newblock {\em IEEE Transactions on Signal Processing}, vol. 63, no. 11, pp.
  2733--2748, 2015.

\bibitem{zhao2012clustering}
X.~Zhao and A.~H. Sayed,
\newblock ``Clustering via diffusion adaptation over networks,''
\newblock in {\em 3rd International Workshop on Cognitive Information
  Processing, 2012. (CIP 2012)}, 2012, pp. 1--6.

\bibitem{zhao2015distributed}
X.~Zhao and A.~H. Sayed,
\newblock ``Distributed clustering and learning over networks,''
\newblock {\em IEEE Transactions on Signal Processing}, vol. 63, no. 13, pp.
  3285--3300, 2015.

\bibitem{cheneusipco2015}
J.~Chen, C.~Richard, and Sayed~A. H.,
\newblock ``{Adaptive clustering for multi-task diffusion networks},''
\newblock in {\em IEEE 23rd European Signal Conference, 2015. EUSIPCO 2015},
  2015, pp. 2746--2750.

\bibitem{khawatmi2015}
S.~Khawatmi, A.~M. Zoubir, and Sayed~A. H.,
\newblock ``{Decentralized clustering over adaptive networks},''
\newblock in {\em IEEE 23rd European Signal Conference, 2015. EUSIPCO 2015},
  2015, pp. 2746--2750.

\bibitem{sayed2012diffusion}
A.~H. Sayed,
\newblock ``Diffusion adaptation over networks,''
\newblock {\em To appear in E-Reference Signal Processing, R. Chellapa and S.
  Theodoridis, Eds., Elsevier, 2013}, 2012 [Online]. Available:
  http://arxiv.org/abs/1205.4220.

\end{thebibliography}

\end{document}